\documentclass[twocolumn]{aastex62}

\received{\today}
\revised{...}
\accepted{...}
\submitjournal{ApJ Letters}

\shorttitle{Star formation from white dwarfs}
\shortauthors{J. Isern}

\usepackage{amssymb}
\usepackage{amsmath}
\usepackage{graphicx}
\usepackage{natbib}
\usepackage{txfonts}
\usepackage{microtype}
\usepackage{subfigure}

\bibpunct{(}{)}{;}{a}{}{,}

\begin{document}

\title{The star formation history in the solar neighborhood as told by massive white dwarfs}

\correspondingauthor{Jordi Isern}
\email{isern@ice.cat, isern@ieec.cat}

\author[0000-0002-0819-9574]{Jordi~Isern}
\affiliation{Institut de Ci\`encies de l'Espai (ICE,CSIC)\\
                 C/Can Magrans, Campus UAB\\
                 08193 Cerdanyola,   Spain}
\affiliation{Institut d'Estudis Espacials de Catalunya (IEEC),\\ 
                 Ed. Nexus-201,  c/Gran Capit\`a 2-4,\\ 
                 08034 Barcelona, Spain}

\begin{abstract}
White dwarfs are the remnants of low and intermediate mass stars. Because of electron degeneracy, their evolution is just a simple gravothermal process of cooling. Recently, thanks to Gaia data, it has been possible to construct the luminosity function of massive ($0.9 \le M/M_\odot \le 1.1$)  white dwarfs in the solar neighborhood ($d < 100$ pc). Since the lifetime of their progenitors is very short, the birth times of both, parents and daughters, are very close and allow to reconstruct the (effective) star formation rate. This rate started growing from zero during the early Galaxy and reached a maximum 6-7 Gyr ago. It declined and $\sim 5$~Gyr ago started to climb once more reaching a maximum 2 - 3 Gyr in the past and decreased since then. There are some traces of a recent star formation burst, but the method used here is not appropriate for recently born white dwarfs. 
\end{abstract}


\section{Introduction} \label{sec:intro}

The luminosity function is defined  as the number  of white dwarfs  of a given  luminosity per
unit volume (or galactic disk surface unit, for instance) and magnitude interval (WDLF from now):

\begin{equation}
N(l) = \int^{M_{\rm s}}_{M_{\rm i}}\,\Phi(M)\,\Psi[T-t_{\rm cool}(l,M)-t_{\rm PS}(M)]
\tau_{\rm cool}(l,M) \;dM
\label{ewdlf}
\end{equation}
\noindent
where  $T$ is the  age of the  population under study,  $l =
-\log  (L/L_\odot)$,  $M$  is  the   mass  of  the  parent  star  (for
convenience all  white dwarfs  are labeled with  the mass of  the main
sequence  progenitor), $t_{\rm  cool}$  is the  cooling  time down  to
luminosity   $l$,    $\tau_{\rm   cool}=dt/dM_{\rm   bol}$    is   the
characteristic cooling time, $M_{\rm s}$ is the maximum mass of a main
sequence star  able to produce a  white dwarf, and $M_{\rm  i}$ is the
minimum mass of the main sequence  stars able to produce a white dwarf
of  luminosity $l$,  i.e.  is  the mass  that satisfies  the condition
$T=t_{\rm cool}(l,M) + t_{\rm PS}(M)$ and $t_{\rm PS}$ is the lifetime
of the progenitor  star.  The  remaining quantities, the
initial  mass  function (IMF, from now),  $\Phi(M)$,  and  the  star  
formation  rate (SFR, from now),
$\Psi(t)$,  are not  known a  priori  and depend  on the  astronomical
properties  of the  stellar population  under study.  Since  the total
density  of white dwarfs  of a  given population  is usually not well
known,  it  is  customary  to  normalize  the  computed  luminosity
function to a bin with a small error bar in order to  compare  
  theoretical  and  observational data. For instance, in the case of the  
disk this bin is usually $l=3$. Therefore, if the
observed luminosity  function and  the evolutionary behavior  of white
dwarfs are known it is possible to obtain information about the properties
of the population under study. 
Evidently, given the nature of the problem, there is always a degeneracy between the galactic properties (SFR and IMF) and the adopted stellar models. 

The process of obtaining such information can be formulated as follows. Let be $t_{\rm b} = T_{disk}  - t_{cool} (l,M) - t_{PS} (M) $ the time at which the progenitor of the white dwarf was born  and  $M=M(t_{\rm b})$ the mass of the star that, being born at this time, is able to  produce a white dwarf  of luminosity $l$ at present. Equation~\ref{ewdlf} can be written as:
\begin{equation} 
 N(l) = \int\limits_0^{t_b^{up} } {K\left( {l,t_b } \right)\Psi \left( {t_b } \right)dt_b } 
\end{equation}
\noindent
with
\begin{equation} 
 K\left( {l,t_b } \right) = \Phi \left[ {M\left( {t_b } \right)} \right]\tau _{cool} \left[ {l,M\left( {t_b } \right)} \right]\frac{{dM\left( {t_b } \right)}}{{dt_b }}
\end{equation}
\noindent
 The kernel, $K(l,t_{\rm b})$, of this integral function is not symmetric in $l$ and $t_{\rm b}$ and it has a quite complicated behavior. Consequently, according the Picard-Lindel\"of's theorem, $\Psi$ cannot be directly obtained and the unicity of the solution is not guaranteed \citep{iser95}.

 One way to tackle the problem is to optimize the parameters of some trial functions comparing, after defining some weight function, models with data \citep{iser99}. Obviously, this solution is optimal within the context of the adopted model, which might not correspond with the reality. Another way consists on, starting from a simple initial guess of the SFR, iteratively improve the solution using all the observational bins until a satisfactory solution is found \citep{rowe13a}. This solution is quite sensitive to the adopted metallicity and IMF, but not to the DA non-DA white dwarf ratio nor the relationship between the mass of the white dwarf and that of the progenitor. All in all, the quality of the final solution essentially depends on the quality of the observational data. 

Finally, if the luminosity function is restricted to  massive white dwarfs the SFR can be directly obtained \citep{diaz94}. This method, however, has suffered from the scarcity of  high mass white dwarfs known. In an early work, this SFR was obtained from the data of \citet{sion88} and \citet{berg92}, and from \citet{legg98} respectively, but the relatively small number of stars in the sample prevented to obtain firm conclusions \citep{iser99}. Fortunately this situation has recently changed thanks to the work of \citet{trem19} who have been able to build a reliable and precise luminosity function of massive stars using the data provided by Gaia.

\section{Massive white dwarfs and the star formation rate}
This luminosity function, averaged over an interval of luminosity $\Delta l$,
can also be directly computed as follows \citep{iser99}. Assume a stellar population that forms at a rate $\Psi(t)$. After a time $T$, the number of white dwarfs that have a luminosity $l$ per unit of luminosity interval is given by
\begin{equation}
N(l,T) = \frac{1}{{\Delta l}}\int\limits_t {\int\limits_M {\Phi \left( M \right)\Psi \left( t \right)dMdt} } 
\end{equation}
where, as before, $M$ is the mass of the parent star, and the integral is constrained to the domain
\begin{equation}
T - {t_{cool}}\left( {M,l - 0.5\Delta l} \right) \le t + {t_{MS}}\left( M \right) \le T - {t_{cool}}\left( {M,l + 0.5\Delta l} \right)
\end{equation}
for all the stars able to produce a white dwarf.

If the integral is restricted to massive white dwarfs, i.e. those for which it is possible to neglect the lifetime of the progenitor in front of the cooling time, and $\Psi (t)$ is smooth enough\footnote{This method is also valid for white dwarfs with masses within a limited enough range of values.}, then

\begin{equation}
N(l,T) \simeq \frac{\left\langle \Psi  \right\rangle}{\Delta l} \int\limits_{\Delta M} {\Phi \left( M \right)\Delta {t_{cool}}\left( {l,M} \right)dM}
\end{equation}
with
\begin{equation}
\Delta {t_{cool}} = {t_{cool}}\left( {l + 0.5\Delta l,M} \right) - {t_{cool}}\left( {l - 0.5\Delta l,M} \right)
\end{equation}
and consequently,
\begin{equation}
\left\langle \Psi  \right\rangle  = \frac{{N(l,T)}\Delta l}{{\int\limits_{\Delta M} {\Phi \left( M \right)\Delta {t_{cool}}\left( {l,M} \right)dM} }}
\label{psi}
\end{equation}

\begin{equation}
\left\langle t \right\rangle  = \frac{{\int\limits_{\Delta M} {\Phi (M)tdM} }}{{\int\limits_{\Delta M} {\Phi (M)dM} }}
\end{equation}

\begin{equation}
\left\langle {\Delta t} \right\rangle  = \frac{{\int\limits_{\Delta M} {\Phi (M)\Delta tdM} }}{{\int\limits_{\Delta M} {\Phi (M)dM} }}
\end{equation}

It is important to notice here that the star formation rate obtained in this way is an effective one in the sense that it recovers the present age distribution of the sample, but does not take into account the secular evolution of the sample mainly due to radial migrations and height inflation. On another hand, hidden WD in binaries and non-resolved double degenerates can bias the sample, and double degenerate mergers can reduce the density of WD in some bins and, in the case they do not explode as SNIa reappear as newly born hot single WD with the corresponding density increase of younger bins, thus modifying the SFR deduced from these data. The importance of this effect is small given the present level of precision, but it will be necessary to include it in order to interpret future high precision WDLFs.

\section{Results and conclusions} 

Table~1 shows the values taken by $\left\langle \Psi  \right\rangle$, $\left\langle t \right\rangle $ and $\left\langle {\Delta t} \right\rangle$ using the \citet{trem19} data and the BaSTI models\footnote{ Cooling models publically available at: \url{http://albione.oa-teramo.inaf.it}.} for DA white dwarfs. Models labeled \emph{ns} take only into account the release of latent heat upon crystallization, while models labeled \emph{s} take also into account  the gravitational energy released by the sedimentation induced  by the changes of solubility during the crystallization process. Both families of models are built with the chemical profiles predicted by the evolution of the progenitor  which depend on the mass \citep{sala10}. The relationship between the masses of the progenitor and white dwarf is that found by \citet{elba18}\footnote{ The results obtained with the \citet{cata08} initial final mass relationship are similar. }, while the IMF is that of Salpeter truncated at 0.1~M$_\odot$ and normalized to the unit mass.  

\begin{table}[h!]

\centering 

\caption
{Total star formation rate $\Psi$~(M$_\odot$Gyr$^{-1}$pc$^{-3}$), age, $t$ and time interval $\Delta t$ (Gyr) obtained from each luminosity function bin. Subindexes s and ns correspond to the cases  with and without sedimentation.}
\begin{tabular}{crrcrrc}

\tablewidth{0pt}
\hline
\hline

{$log_{10}(L/L_\odot)$} &

{$t_{\rm s}$} &

{$\Delta t_{\rm s}$} & 

{$\log_{\rm 10} \Psi_{\rm s}$} &

{$t_{\rm ns}$} &

{$\Delta t_{\rm ns}$} & 

{$\log_{\rm 10} \Psi_{\rm ns}$} \\

\hline
  -1.20  &   0.05  &   0.04  & -2.794  &   0.05  &   0.04  & -2.794 \\
  -1.70  &   0.12  &   0.16  & -2.553  &   0.12  &   0.16  & -2.553 \\
  -2.30  &   0.41  &   0.43  & -2.655  &   0.41  &   0.42  & -2.643 \\
  -2.80  &   0.97  &   0.81  & -2.780  &   0.91  &   0.64  & -2.678 \\
  -3.10  &   1.80  &   0.70  & -2.546  &   1.53  &   0.52  & -2.418 \\
  -3.30  &   2.59  &   0.88  & -2.468  &   2.13  &   0.67  & -2.350 \\
  -3.50  &   3.53  &   0.99  & -2.600  &   2.86  &   0.80  & -2.508 \\
  -3.70  &   4.58  &   1.11  & -2.747  &   3.75  &   0.98  & -2.694 \\
  -3.90  &   5.75  &   1.23  & -2.753  &   4.82  &   1.16  & -2.728 \\
  -4.10  &   7.06  &   1.46  & -2.667  &   6.09  &   1.43  & -2.660 \\
  -4.30  &   8.88  &   2.58  & -2.885  &   7.89  &   2.57  & -2.884 \\
  -4.50  &  11.95  &   2.82  & -3.403  &  10.96  &   2.82  & -3.403 \\
  -4.70  &  14.13  &   1.66  & -4.123  &  13.14  &   1.68  & -4.130 \\
\hline

\end{tabular}

\end{table}

\begin{figure}[h!]
\centering
\includegraphics[width=9.0cm]{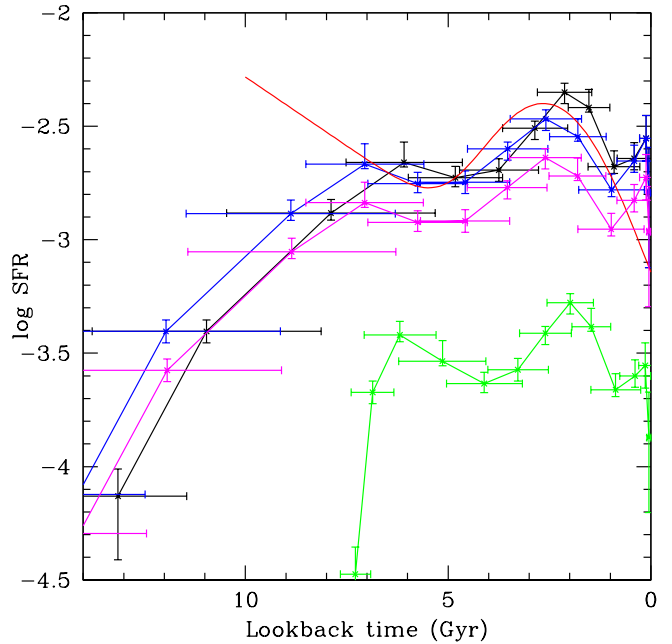}
\caption{Star formation rate (M$_\odot$ Gyr$^{-1}$pc$^{-3}$) in the solar neighborhood  obtained from massive white dwarfs ($d\le 100$ pc). Blue dots were computed taking into account the energy released by crystallization (latent heat) and induced sedimentation, and black ones only latent heat. The red line represents the star formation rate per unit of disk surface obtained from main sequence stars \citep{mor19}, divided by and arbitrary scale height to allow comparisons. Magenta points were computed in the same way as the blue ones but using the IMF of \citet{mor19}. Green points represent the SFR, divided by 10 just for a sake of clarity, obtained with the Montreal models.}  
\label{fig1}
\end{figure}

Figure~1 displays these results, where blue and black dots correspond to the calculations with and without sedimentation respectively. As it can be seen, in both cases the effective star formation rate is not a monotonically decreasing or constant function as it is often assumed. It grew quickly in the past, during the first epochs of the Galaxy, and roughly stabilized and started to decrease  at 7 or 6  Gyr ago (cases s and ns respectively) around the values $\log_{\rm 10} \Psi \approx -2.4, -2.8$ M$_\odot$Gyr$^{-1}$pc$^{-3}$. A noticeable feature is the prominent peak centered at 2.8 or 2.2  Gyr ago depending on the adopted cooling model. The increase of the SFR near the present time is not reliable since it does not satisfy the hypothesis of a negligible main sequence lifetime versus cooling time and deserves more attention. 

A hint of this behavior, a bump centered around 2-3 Gyr, was already present in the results obtained by \citet{iser99} --see their  Figure 2--  but it was not interpreted as indicative of star formation variability.  The small number of stars in the sample prevented its identification, in contrast with the present situation, where the quality of the  \citet{trem19} luminosity function provides a robust argument in favor of a non monotonous behavior of the SFR.

Interestingly, \citet{rowe13a,rowe13b} inverted the total luminosity functions obtained by \citet{harr06} and \citet{rowe11} from the Sloan and the SuperCOSMOS Sky Surveys respectively and found in both cases a solution characterized by two peaks of star formation, placed at  $\sim 9$ and $2 - 3 $~Gyr in the past, in qualitative agreement with the results found here.

The existing degeneracy between galactic properties and evolutionary models implies that different models can lead to different star formation histories. The green dots of Fig.~1 display the evolution of the SFR obtained with the Monreal 
models\footnote{\url{http://www.astro.umontreal.ca/~bergeron/CoolingModels}} COXXX0210 which are made of a half oxygen half carbon core, a He-layer of $10^{-2}$~M$_\odot$ and a H-layer of $10^{-10}$~M$_\odot$ and do not take into account sedimentation. In this case the bump is present, but the star formation abruptly starts around $\sim 7$~Gyr.

One way to remove this degeneracy is to compare these results with other star formation histories that have been obtained with independent methods.
The red line of Fig.~1 displays, after dividing by an arbitrary scale height to allow comparisons, the SFR per unit of galactic surface disc obtained with the Gaia DR2 data for Main Sequence stars with $G \le 12$ in the context of the Besan\c{c}on Galaxy Model \citep{mor19}. This analysis suggests a decreasing trend in the interval of 9-10 to 6-7 Gyr followed by a star burst with a maximum centered at 2-3 Gyr.
Magenta dots were obtained as in the sedimentation case but adopting the IMF proposed by \citet{mor19} in their analysis of the Gaia data. The similarity of both computed SFRs is due to the fact that this IMF is not too different from the Salpeter's one in the region corresponding to the masses of the progenitors of the massive white dwarfs considered here. Two facts deserve attention. i) the position and the width of the SFR burst obtained by \citet{mor19} seems to favor models including sedimentation, and ii) the local and the disc SFR seem to diverge at the early epochs of the Galaxy.
This last behavior can have several origins and demands further attention. One possibility is a delay in starting the star formation process respect to inner regions of the disc \citep{kubr15} or just a different behavior of the outer disc as compared with the inner one, as proposed  by \citet{hayw18}. Another one is a vertical dilution induced by a galactic collision like the Gaia-Enceladus event \citep{helm18}.
 
\begin{figure}
\centering
\includegraphics[width=9.0cm]{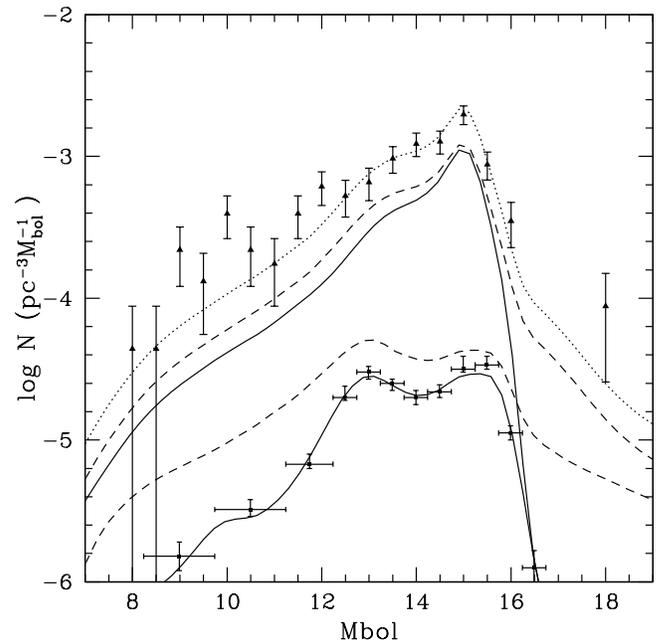}
\caption{Theoretical luminosity functions obtained from the SFR of Table ~1 (case \emph{s}).  Solid lines: massive (bottom) and all masses (top) DAs, excepting ONe ones. Dashed lines: all massive white dwarfs (DAs and non-DAs and ONe ones (bottom) and all DAs (top). Dotted line: the same as the top dashed line, but normalized to the total luminosity function. Squares: DA white dwarfs of all masses excepting ONe ones  \citet{trem19}. Triangles: all white dwarfs \citet{oswa17}.  
 }
\label{fig2}
\end{figure}

Since the SFR has been derived from the tail of the mass distribution of white dwarfs and neglecting the lifetime of the progenitor, it is natural to check if it can reproduce the luminosity function of all white dwarfs in the solar vicinity. For that purpose, Fig.~2 displays a LF that is representative of all white dwarfs present in a volume of 25~pc around the Sun and it is believed to be 68\% complete \citep{oswa17}. Figure 2 also displays the luminosity function of massive DA white dwarfs and the corresponding theoretical counterpart (solid line). The dashed line is obtained when non-DAs and white dwarfs with massive ONe cores are included. The total luminosity function is represented by black lines (dashed for all white dwarfs, solid for DAs with CO cores only). As it can be seen the shape is well reproduced except for a peak at $M_{\rm bol}\sim 9$, which can be accounted for placing a burst at $\sim 0.4$~Gyr, which is in the limit of the method presented in this Letter. A potential problem is that the total WDLF predicted with the SFR obtained here is a factor $\sim 3$ smaller than the observed one. 

The uncertainties in the IMF and in the initial-final mass relationship, as well as the way as the scale height over the galactic plane is included, alleviate the discrepancy but does not solve it. Other possibilities are the degree of completeness of the solar sample or the secular galactic evolution in the solar neighborhood, but given the present uncertainties it is not possible to obtain any definite conclusion and it will be necessary to wait for a distribution not only in luminosities but also in masses. 

Just to conclude it can be said that massive white dwarfs provide a robust argument in favor of a  star formation burst in the solar neighborhood that occurred 2-3 Gyr ago as well a a hint of the existence of a more recent one, around 0.4-0.3 Gyr. These results are a clear demonstration of the possibilities offered by white dwarf cosmochronology to study the evolution of the Galaxy and the necessity to completely understand their physical properties.

\acknowledgements
This work  has been supported  by MINECO grant  ESP2017-82674-R, 
by EU FEDER funds, and by
grants 2014SGR1458 and CERCA Programe of  the Generalitat de
Catalunya.


\begin{thebibliography}{}
\bibitem[Bergeron, Saffer \& Liebert(1992)]{berg92}Bergeron, P., Saffer, R.~A. \& Liebert, J \ 1992, \apj, 394,228
\bibitem[Catalan et al.(2008)]{cata08}Catal{\'a}n, S., Isern, J., Garc{\'{\i}}a-Berro, E. \& Ribas, I. \ 2008, \mnras, 387, 1693
\bibitem[Diaz-Pinto et al.(1994)]{diaz94}Diaz-Pinto, A., Garc{\'{\i}}a-Berro, E., Hernanz, M., Isern, J. \& Mochkovitch, R. \ 1994, \aa 282, 86
\bibitem[El-Badry et al.(2018)]{elba18}El-Badry, K., Rix, H-W, \& Weisz, D.~R.\ 2018, \apjl, 860, L17
\bibitem[Harris et al.(2006)]{harr06}Harris, H.C., Munn, J.C.,  Kilic, M. et al. \ 2006, \aj 131, 571
\bibitem[Haywood et al.(2018)]{hayw18}Haywood, M., Di Matteo, P., Lehnert, M. et al. \ 2018, \aap, 618, A78
\bibitem[Helmi et al.(2018)]{helm18}Helmi, A., Babusiaux, C., Koppelman, H.~H. et al. \ 2018, \nat, 563, 85
\bibitem[Isern et al.(1995)]{iser95}Isern, J., Garc{\'{\i}}a-Berro, E., Hernanz, M., Mochkovitch, R., \& Burkert, A.\ 1995, in White Dwarfs, Ed. D. Koester \& K. Werner, Lecture Notes in Physics, 443, 19
\bibitem[Isern et al.(1999)]{iser99}Isern, J., Hernanz, M., Garc{\'{\i}}a-Berro, E., \& Mochkovitch, R. \ 1999, in White Dwarfs, Ed. S.~E. Solheim \& E.~G. Meistas, ASPCS 169, 408 
\bibitem[Kubryc et al.(2015)]{kubr15}Kubryk, M., Prantzos, N. \& Athanassoula, E. \ 2015, \aap, 580, A126
\bibitem[Legget et al.(1998)]{legg98}Leggett, S.~K., Ruiz, M.~T., Bergeron, P. \ 1998, \apj, 497, 294
\bibitem[Mor et al.(2019)]{mor19}Mor, R., Robin, A.~C., Figueras, F., Roca-F{\`a}brega, S. \& Luri, X. \ 2019, ArXiv 1901.07564
\bibitem[Oswalt et al.(2017)]{oswa17}Oswalt, T.~D. Holberg, J. \& Sion, E. \ 2017, in 20th European White Dwarf Workshop, Ed. P. E. Tremblay et al., ASP Conference Series 509, 59
\bibitem[Rowell(2013a)]{rowe13a} Rowell, N. \ 2013a, \mnras, 434, 1549
\bibitem[Rowell(2013b)]{rowe13b} Rowell, N. \ 2013b, in 18th European White Dwarf Workshop, Ed. Krzesinsky et al., ASP Conference Series  469, 89
\bibitem[Rowell \& Hambly(2011)]{rowe11}Rowell, N. \& Hambly. N.C. \ 2011, \mnras, 417, 93
\bibitem[Salaris et al.(2010)]{sala10}Salaris, M., Cassisi, S., Pietrinferni, A., Kowalski, P.~M. \& 
Isern, J \ 2019, \apj, 716, 1241
\bibitem[Sion et al.(1988)]{sion88}Sion, E.~M., Fritz, M.~L., McMullin, J.~P. \& Lallo, M.~D. \ 1988, \aj, 96, 251
\bibitem[Tremblay et al.(2019)]{trem19}Tremblay, P.-E., Fontaine, G., Fusillo, N.~P.~G., Dunlap, B.~H., et al. \ 2019, Nature, 565, 202


\end{thebibliography}
\end{document}